\documentclass[preprint,aps]{revtex4}
\begin{document}
\title{A Critical String Theory in 3+1 Dimensions}
\author{J.S. Bhattacharyya}
\altaffiliation[Permanently at ]{Physics Department, Kanchrapara College, Kanchrapara, W.B. India.}
\author{Gautam Bhattacharya}
\email{gautam@theory.saha.ernet.in}
\affiliation{Theory Division, Saha Institute of Nuclear Physics,\\ 1/AF Bidhannagar, Kolkata 700064, India}

\begin{abstract}
Redefining the vacuum state of a free twofold $N=1$ covariant supersymmetric string action as the one with all the world sheet fermionic excited states occupied, makes the theory anomaly free in $D=4$ with Minkowski signature. The theory thus describes a critical string in $3+1$ dimensions as opposed to earlier $N=2$ supersymmetric  theories describing a $2+2$ dimensional target space. While in the NS sector the spectrum  basically resembles the same for the standard $N=1$ superstring theory with one of the $N$ species in the background, in the R sector both the species of fermions and superconformal ghosts are required to describe the relevant spin operators to describe the fermion spectra. 
A crucial difference from $D=10$ case is that the fermion states are Dirac particles instead of Majorana-Weyl. Even though the full spectrum of the theory contains both bosons and fermions of various spin, there is no space-time supersymmetry due to obvious lack of triality.
\end{abstract}
\maketitle
Even though the studies of critical strings in 26 (for bosonic cases) and 10 (for $N=1$ supersymmetric cases) space-time dimensions  opened up an exciting new world in theoretical physics and mathematics, it has always remained a distant dream to have one such theory manifestly critical in $3+1$ dimensions as that could then be used to describe the real world physics like the non-perturbative aspects of QCD and quantum theory of Einstein gravity. Several years back  Ooguri and Vafa \cite{vafa1} went tantalizingly close to it when they showed that the canonical $N=2$ supersymmetric string can be made to have a four dimensional target space where the two extra dimensions would be associated with Liouville and the $SO(2)$ gauge anomaly  modes. Unfortunately the metric turned out to have a non-Minkowski signature and therefore untenable to describe physics in the real world.  What we find to our pleasant surprise is that a simple redefinition of
the ground state of a two-fold $N=1$ superstring (It mimics  $N=2$ supersymmetric string in many respects) solves the problem. A simple computation of the terms proportional to $m^3$ of the Virasoro anomaly $A(m)$ from the constituent fields in the $N=2$ supersymmetric case without the additional local symmetries like $SO(2)$ will be the sum of $2D$ , $D$, $+22$ and $-26$  respectively from the two species of bosons, fermions, superconformal ghosts and a single species of the conformal ghosts. Unfortunately this  vanishes for  $D=\frac{4}{3}$ and would be physically inconsequential. However inclusion of other ghost terms in the action arising from gauge fixing that eliminates the $SO(2)$ gauge field will make the theory anomaly free for integer value of  $D$ but with value of 2 \cite{GSW1} which is not so interesting from the point of view of string physics.  Quite remarkably  we find that with a changed definition of the world sheet fermionic vacuum even without the local $SO(2)$ symmetry leads to an anomaly free solution for integer $D$ and the value turns out to be indeed 4.  The additional advantage of keeping the symmetry global  is that there will not be any transition from one species of $N=2$ to the other -- rendering one species in the background while considering the dynamics of the other. In that sense it is effectively a $N=1$ superstring theory in a $3+1$ dimensions. While constructing the vertex operator for fermion emission in the Ramond (R) sector, however, there will be a need for mixing of the two species of fermions and the superconformal ghosts to define the suitable spin operators in $D=4$.  Only there  the second species  play a role in a subtle way.

We demonstrate the cancellation of the anomalies by invoking the well known textbook methodology \cite{GSW1} of starting with the world sheet action involving free bosons and fermions along with the conformal and superconformal ghost fields. Making the bosonic and fermionic fields two-fold would necessitate incorporation of two-fold superconformal ghosts during  gauge fixing to eliminate the gravitino fields but the conformal ghosts will remain the same as before. Without the ghost terms (which are essential for quantum theory) the action mimics a gauge fixed $N=2$ world sheet supersymmetric theory, though we will prefer to call it a two fold $N=1$ superstring theory.
Each one of these fields will have Fourier expansion with the modes appropriately chosen to match the boundary conditions. The generators of the Virasoro algebra are the bilinears of these Fourier modes. The bilinears are normal ordered so as not to give divergent results with respect to the canonical vacuum which is defined as the one annihilated by all positive Fourier modes.  However  in case of fermions there can be another choice of a vacuum where all energy levels are filled up with exactly one fermion per level according to Pauli principle. This is a unique state too and all states built from this would only contain excitations of the bosonic field oscillators and/or dexcitations (obtained by action of positive fourier modes) of the fermion. The anomaly term of the fermionic part of the Virasoro algebra, calculated based on such infinite Fermi sea, will be shown to have the same value but with opposite sign compared to the  one obtained from the canonical Fock space vacuum. This is because, as is well known, the anomaly term involves only the odd powers of the Fourier indices ($m$ and $m^3$) and the redefinition of vacuum effectively reverses the indices (role reversal of the creation and destruction operators).  The scenario is somewhat analogous to Dirac's original hole theory \cite{Dirac} where the ground state was defined to be the one with all negative energy states filled up.

One might wonder how the anomaly, which is basically a c-number central extension of an  operator algebra, would have a different form with what looks like a partly  upside down situation of the original Fock space (i.e. only as far as the fermions are concerned). The reason behind such difference is the fact that the normalization coefficient of the new vacuum is infinite with respect to the canonical one (zero boson and zero fermion state) because of infinite number of fermions that defines the former. One should note here that the $N=2$ world sheet supersymmetry which is manifest in the action is not quite reflected in the usual fashion in the states. The bosonic state created by the creation operator of a boson field on the vacuum will transform into a ``hole'' state of the fermions obtained by the action of the destruction operator of the fermion on the Fermi vacuum. In that sense it can be termed as {\em antisupersymmetry}.

While for bosons for each species we take the usual Fourier expansion  
\begin{equation}
X^\mu = \displaystyle {x^\mu+p^\mu\tau+i\sum_{n\neq 0} \frac{1}{n}\alpha^\mu_ne^{-in\tau}\cos n\sigma}
\end{equation}
For fermions we take, as a specific case, the open string boundary condition of Neveu-Schwaz(NS) type for each species 
\begin{equation}
\psi^\mu_+ = \displaystyle{\frac{1}{\sqrt2}\sum_{r\in Z+1/2}
b^\mu_re^{-ir(\tau+\sigma)}} \label{NS}
\end{equation}

For each of the two species of  the superconformal ghost the Fourier expansion can be written as
\begin{equation}
\begin{array}{l}
\gamma_{1/2}=\displaystyle{\frac{1}{\sqrt2}\sum_{r\in Z+1/2}\gamma_re^{-2ir(\tau+\sigma)}}\\
\beta_{-3/2}=\displaystyle{\frac{1}{\sqrt2}\sum_{r\in Z}\beta_re^{-2ir(\tau+\sigma)}}
\end{array}
\end{equation}
and that for the single species of the conformal ghost is
\begin{equation}
\begin{array}{l}
c^+ = \displaystyle{\sum_{n=-\infty}^\infty c_ne^{-in(\tau+\sigma)}}\\
b_{++} = \displaystyle{\sum_{n=-\infty}^\infty b_ne^{-in(\tau+\sigma)}}\\
\end{array}
\end{equation}

The contributions from the two species of the bosonic field $X^\mu$ and the single species of the conformal ghosts are well known \cite{GSW1}, namely
\begin{equation}
A^\alpha(m)=\frac{D}{6}(m^3-m)\label{alpha}
\end{equation}
\begin{equation}
A^c(m)=\frac{1}{6}(m-13m^3)\label{conformal}
\end{equation}

What turns out to be drastically new is the central extension of the algebra of the generators $L^b_m$ defined through
$$ L^b_m=\frac{1}{2}\sum_{-\infty}^\infty (r+\frac{m}{2}):b_{-r}.b_{m+r}:$$
with $b$'s are as defined in Eq.(\ref{NS}).  First we notice that
$$
[L^b_1,L^b_{-1}]=\frac{1}{4}\sum_{-\infty}^\infty
(r-1/2)(r^\prime+1/2)[b_r.b_{1-r},b_{r^\prime}.b_{-1-r^\prime}]$$
When one uses the fact that all creation operators ($b_r$ with $r<0$) will destroy the ground state which are already saturated with fermions and physical states are the ones that satisfy the Virasoro conditions, 
the right hand side vanishes identically. 
Similarly starting with the $L_2$'s we have
$$[L^b_2,L^b_{-2}]=\frac{1}{4}\sum(r-1)(r^\prime+1)
[b^\mu_r b^\mu_{2-r},b^\nu_{r^\prime} b^\nu_{-2-r^\prime}]$$
which, once again, when sandwiched between the physical states based on the redefined ground state and Virasoro condition,
gives
\begin{equation}
<[L^b_2,L^b_{-2}]>=-\frac{D}{4}
\end{equation}
Hence, 
\begin{equation}
A^b(1)=0, \ \ A^b(2)=-D/4. \label{A1A2}
\end{equation}
for each species of the fermion.
 Since  Jacobi identity implies that an arbitrary $A^b(m)$ has the general form
$A^b(m)=c^bm^3+d^bm$, 
Eq.(\ref{A1A2}) implies
\begin{equation}
A^b(m)=-\frac{D}{24}(m^3-m)\label{fermion}
\end{equation}
Notice the crucial negative sign before $m^3$ which was absent when the same computation was done with respect to
the canonical vacuum.

The central extension of the Virasoro algebra of the generators corresponding to the superconformal ghosts defined by
\begin{equation}
L^\gamma_m=\sum_{-\infty}^\infty (r+\frac{m}{2}):\beta_{m-r}\gamma_r:
\end{equation}
with the usual definition of normal ordering comes out to be
\begin{equation}
A^\gamma(m)=\frac{1}{12}(11m^3+m)\label{superc}\label{sign}
\end{equation}
The positive sign for the linear power in $m$ is an artifact of the normal order definition. We could, keeping similarity with the fermions and the new definition of vacuum, define the normal ordering of the ghosts to be one with all destruction operators to the left of creation operators. This will only add a Zeta-function regularized constant to the  anomaly term in Eq.(\ref{sign}) and would eventually change the sign of the coefficient of $m$.

The net contribution from the two species of the superconformal ghosts will, therefore, be twice  of this. The same must also be the case for the fermions as well.

Noting that the cancellation of anomaly is primarily related to $m^3$ while the coefficient of $m$ is always adjustable by shifting of the vaccum,  we just add up the individual contributions proportional to the $m^3$ term and that turns out to be
$\frac{1}{12}(D-4)$, showing clearly the vanishing of anomalies and hence the criticality at $D=4$. This is the main result of the paper. 

As for the terms proportional to $m$ in the anomaly, we notice that
since any constant $a^\prime$  can be added
to the $L_0$ 
the anomaly term $A(m)$ defined through the centrally extended Virasoro algebra
\begin{equation}
[L_m,L_n]=(m-n)L_{m+n}+A(m)\delta_{m+n}
\end{equation}
can now be written as:
\begin{equation}
A(m)=A^\alpha(m)+A^b(m)+A^c(m)+A^\gamma(m)+2a^\prime m \label{anomaly}
\end{equation}
Substituting the values obtained in Eqs.(\ref{superc},\ref{alpha},\ref{conformal},\ref{fermion}) one gets
\begin{equation}
A(m)=\frac{m^3}{12}(D-4)-\frac{m}{12}(D-24a^\prime)
\end{equation}
Thus the complete anomaly $A(m)=0$ iff $D=4$ and $a^\prime=1/6$.

Similarly using the other boundary condition (R sector) where the Fourier coefficients of $\psi$ are $d^\mu_n (n\in Z)$, but maintaining the same boundary condition and interpretation for the superconformal ghosts as before (this would be consistent with the spinor vertex operators to be discussed shortly afterwards in the context of vertex operators),  one gets,
\begin{equation}
A(m)=\frac{m^3}{12}(D-4)-\frac{m}{12}(4D-24a^\prime)
\end{equation}
Once again the theory is anomaly free for $D=4$ and $a^\prime=2/3$.

The different values of $a^\prime$ namely $1/6$ and $2/3$ for the NS and R sector respectively are, however, not to be taken in their face values for the actual string amplitude calculations. Recall that the action of $L_0^b$ and $L_0^d$ which is to be identified with the string Hamiltonian for NS and R sectors, have  nontrivial action on the newly defined vacuum. This is because the operator form of $L_0$ contains for each mode a fermion number operator ($b_{-r}.b_{r}$ or
$d_{-n}.d_{n}$ $r,n >0$) whose action on physical state is unity term by term. Hence on ground state the action of $L_0^b$
will produce the result
$$L_0^b|0> =\sum_{r\in Z+1/2 >0}r b_{-r}.b_r|0> =\sum r|0>= \frac{2}{6}|0> $$
Similarly the action of $L_0^d$ would give
$$L_0^d|0> =\sum_{n\in Z >0}n d_{-n}.d_n|0>=\sum n|0> = -\frac{2}{3}|0> $$
Here we have used the standard $\zeta$-fn regularization and doubling of fermions for $N=2$.
These background constants are to be absorbed  in $a^\prime$ defined earlier separately in NS and R sectors to define the actual shift of the ground state mass  and to identify the string Hamiltonian with $L_0$. In other words we see that for physical states  the value of $L_0$ is $\frac{1}{2}$ and zero for NS and R sectors respectively (exactly the same as what one has in the canonical theory of $N=1$ supersymmetric string in the respective sectors). It is obvious that identical results would be obtained for closed strings with both NS and R sectors for left and right moving modes independently.

It should also be pointed out that Jacobi identity ensures  the worldsheet supersymmetry  at the quantum level too.

Having obtained an anomaly free  superstring theory describing a dynamics in 3+1 dimensions,
the next step is to get the spectra of the physical on-shell states  using the vertex operators for both NS and R sectors. 

\noindent{\em NS sector:}

Recalling that the $\tau$ evolution is defined by the Hamiltonian $L_0$, a typical boson emission/absorption vertex operator $V(\tau)$ for an open string will have the form
$$V(\tau)=e^{iL_0\tau}V(0)e^{-iL_0\tau}$$
with 
$$V(0)=[G_r,W(0)]_\pm$$
where $G_r$ are the Fourier modes of the supercurrent in the NS sector.

The simplest situation is the one when $W =e^{ik.X(0)}$ which would correspond to the tachyon emission. The corresponding
vertex operator $V(0)$ will have the form

$$V(0)=k.\psi e^{ik.X}$$

Such tachyon states can be eliminated by the usual GSO projection \cite{GSO}. 

The remaining scenario for all other particle states would be almost similar to the ones already well known in the $N=1$ superstring theory.

\noindent{\em R sector:}

Here the vertex operator  for the emission of the lowest excitation would be obtained in terms of the spin operators $D_{\pm 1/2}$ and the conformal dimension restoring ghost spin operators $\Sigma_{\pm 1/2}$ \cite{FMS}\cite{KNZ}. Even though there are four different fermionic spin operators corresponding to two species and $D=4$, the states created by the longitudinal and time-like zero-modes would be either unphysical (this happens when they operate singly on a physical ground state where it violates the $F_0$ super-Virasoro condition) or  expressable in terms of the bilinears of the transverse zero modes (this happens when the product of  time-like and longitudinal zero modes operate on the physical ground state, since $\gamma_0\gamma_3=\gamma_5\gamma_1\gamma_2$ and the states, being massless, are eigenstates of $\gamma_5$). In short there are only four allowed lowest lying states which can be achieved by the application of the spin operators and hence one requires a product of the form $D_{\pm 1/2}D_{\pm 1/2}$ in the construction of relevant fermionic vertex operator. A simple way to do this is to bosonize $\psi^1 \pm \psi^2$ associated with the transverse directions $1$ and $2$  for each species, define the corresponding spin operator $D^N_{\pm 1/2}$, ($N=1,2$) and take the product $D^1D^2$. For the superconformal ghosts the spin operators $\Sigma_{\pm 1/2}$ are obtained from the canonical bosonization (in terms of usual massless bosons)  of the complex combinations $\gamma_1\pm i\gamma_2$ from the two species. They can be made to commute (rather than anticommute) by multiplication of  global anticommuting ``spurion'' operators.  Thus one will have a four index vertex operator
\begin{eqnarray}
&&V^\alpha=\Sigma_{1/2}\Sigma_{-1/2}\bar{u}^\alpha\gamma_\mu \theta_\alpha\\\nonumber
&&(\dot{X}^\mu+i\psi^\mu\psi.k)e^{ik.X} \ \ ({\rm no \ summation \ over \ }  \alpha)
\end{eqnarray}
Or
\begin{equation}
W^\alpha=\Sigma_{+1/2}\Sigma_{-1/2}\bar{u}^\alpha\gamma_\mu \theta_\alpha\psi^\mu e^{ik.X}
\end{equation}
where
$\theta_\alpha=D^1_{\pm 1/2}D^2_{\pm 1/2}$. Note that $V$ has the desired conformal dimension $(3/8-5/8+2/8)+1=1$ for $k^2=0$. The asymptotic four component fermion state obtained this way is
\begin{eqnarray}
&&|u^\alpha>=\lim_{\tau\rightarrow i\infty}W^\alpha|0>_B\\\nonumber
&&=|0,k>_\alpha u^\alpha
\end{eqnarray}
Notice here that both the species of $\psi$ contributes to make the spinor state four-component.  It is here that the two-fold concept plays an important role apart from the cancellation of the anomaly discussed earlier. Even though in the NS sector  the second species are always in the background, in the R sector both the species play a role in a mixed fashion through the spin operators for the production of all space-time fermionic states. It should also be noted that the product $\Sigma_{+1/2}\Sigma_{-1/2}$ changes the boundary conditions for the superconformal ghosts twice to restore them to the desired bosonic branch (half-integral modes).

We should point out some of the salient features of the theory developed in this article.

\noindent 1) In order to use the techniques of conformal invariant field theory in path integral approach \cite{Polyakov}
it is necessary to get a proper form of the world sheet fermion propagators for the loop effects in the anomalies. Notice however that under the changed definition of the vacuum, the naive vacuum expectation value of the fermions from its Fourier expansion would lead to
\begin{equation}
<\psi^\mu(y_i)\psi^\nu(y_j)>=\eta^{\mu\nu}\sum_{m,r>0}(\frac{y_i}{y_j})^{m,r}
\end{equation}
where $y=e^\tau$ for imaginary $\tau$ (Euclidean space).

This expression does not converge for  $\tau_i> \tau_j$. To get around it one needs to reparametrize $\sigma^\alpha\rightarrow-\sigma^\alpha$ in the string action. Since this reverses the sign of the kinetic energy term in the action, it would bring in a change of sign of
the fermionic propagator and the corresponding contribution to the Virasoro anomaly. 

\noindent 2) Because of the negative contribution to the Virasoro anomaly from the fermions there is no question of heterosis. So gauge symmetry can only be introduced by the Chan-Paton method involving open strings only.  It should also be noted that since there is no Weyl Projection there  is no chiral anomaly unlike the case in $D=10$ \cite{GSW2}. The theory contains massless four component Dirac particles as has been explicitly shown by the vertex operator in the R sector. More interestingly, because of the lack of triality in four dimensions and the fermions being of Dirac type, there is no space-time supersymmetry in this new theory. 

\noindent 3) One can use the path integral formalism using the opposite signed Fermion propagator to recover Einstein gravity as a consequence of the conformal invariance of the action for the associated nonlinear $\sigma$-model in the critical dimension four directly.

It will be interesting to study, since there is no anomaly, the restoration of unitarity of the Veneziano-Virasaro-Sapiro duality amplitudes in 3+1 dimensions obtained from this new string theory. This will be discussed in a subsequent paper.

\noindent {\em Note added:} It has been brought to our attention that a renormalisation group approach \cite{Alexandre} to the leading order in the Regge slope parameter for a bosonic string coupled in a time dependendent way to a dilaton background can be made consistent in a $3+1$ dimensional target space. It would be interesting to explore whether our free string theory with manifest space-time covariance can be related to this in a specific frame after integrating out all background fields as well as the fermions.


\begin{thebibliography}{10}
\bibitem{vafa1} Ooguri H. and Vafa C.  \newblock{Nucl.
 Phys. {\bf B361}, p469, 1991}
\bibitem{GSW1} Green M.B., Schwartz J.H. and Witten E,  \newblock{Superstring Theory,Vol.1, Cambridge University Press,
1987}
\bibitem{Dirac} Dirac P.A.M. \newblock{Principles of Quantum Mechanics, Cambridge University press, 1935}
\bibitem{Polyakov} B. Polyakov \newblock{Phys. Lett. {\bf 103B}, p207,211 (1981)}
\bibitem{FMS} Friedan D., Martinec E. and Shenker S.,\newblock{Nucl.
 Phys. {\bf B271}, p93, 1986}
\bibitem{KNZ} Knizhnik V.G. \newblock{Phys. Lett. {\bf 160B}, p403 (1985)}
\bibitem{GSO} Gliozzi F., Scherk J. and Olive D. \newblock{Phys. Lett. {\bf 65B}, p282 (1976), Nucl. Phys. {\bf B122}, p253 (1977)}
\bibitem{GSW2} Green M.B., Schwartz J.H. and Witten E.,  \newblock{Superstring Theory,Vol.2, Cambridge University Press,
1987, chapter 10}
\bibitem{Alexandre} Alexandre J., Ellis J., Mavromatos N.E., \newblock{JHEP 0612:071 (2006)}, Alexandre J., Mavromatos N.E., \newblock{arXiv:hep-th/0703171}
\end{thebibliography}
\end{document}